\documentclass[aps,twocolumn,showpacs]{revtex4}
\usepackage{graphicx}
\usepackage{amsmath}
\usepackage{amssymb}
\usepackage{hyperref}
\begin{document}
\title{Critical percolation in self-organized media:\\
A case study on random directed networks}
\author{Christel Kamp}
\email{kamp@theo-physik.uni-kiel.de}
\affiliation{Institut f\"ur Theoretische Physik, Universit\"at Kiel,
Leibnizstr.\ 15, D-24098 Kiel, Germany}
\author{Stefan Bornholdt}
\affiliation{Institut f\"ur Theoretische Physik, Universit\"at Kiel,
Leibnizstr.\ 15, D-24098 Kiel, Germany}
\affiliation{Interdisziplin\"ares Zentrum f\"ur Bioinformatik,
Universit\"at Leipzig, Kreuzstr.\ 7b, D-04103 Leipzig, Germany}

\begin{abstract}
A minimal model for self-organized critical percolation on directed graphs
with activating and de-activating links is studied. Unlike classical
self-organized criticality, the variables that determine criticality are
separated from the dynamical variables of the system and evolve on a
slower timescale, resulting in robust criticality. While activity of nodes
percolates across the network, the network self-organizes through local
adjustment of links according to the criterion that a link's adjacent 
nodes' average
activities become similar. As a result, the network self-organizes to the
percolation transition with activity avalanches propagating marginally
across the graph. No fine-tuning of parameters is needed.
\end{abstract}
\maketitle

\begin{figure}[h]
\centerline{
\includegraphics[width=7cm]{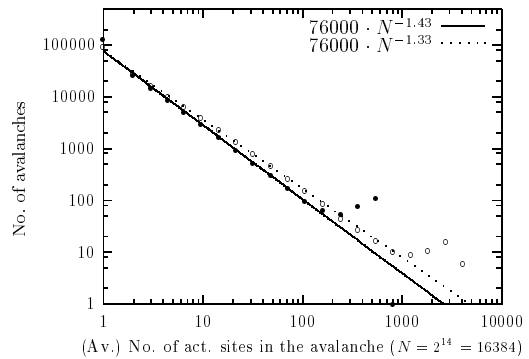}}
\caption{\label{bothavlb} Distribution of avalanche sizes in the
self-organized network, corresponding to the time averaged number
of active sites after equilibration ($\bullet$) and the number of sites
that have ever been activated during the spread of an avalanche ($\circ$).
}
\end{figure}

One astonishing property of many complex natural systems
is their extreme robustness and their stability in the face of a changing
environment. Complex information processing in biological systems,
as neural networks, molecular signaling of the cell, or the immune system,
seems to avoid regimes of chaotic dynamics, which easily could occur
wherever a large number of interacting regulators or switches are
connected. Mechanisms that in principle can stabilize complex interacting
systems have been studied recently in model systems of theoretical
physics. One mechanism that allows a system to self-organize into
a specific global state on the basis of locally acting forces is
Self-Organized Criticality (SOC), enabling dissipative systems
to drive themselves into critical states without any parameter tuning.
Examples are sandpile models  \cite{bak:wiesenfeld:1987} or models
of evolving species relationships in biological evolution
\cite{bak:sneppen:1993}. Similar systems that self-organize to specific
critical states are forest fire models \cite{drossel:schwabl:1992},
and earthquake models \cite{olami:christensen:1992}.
\newline
It is an interesting question whether similarly simple principles of
self-organization are at work in biological information processing systems.
In such systems, often organized as networks of cells or molecules,
the architecture of interactions may contribute to the overall robustness.
Classical models of self-organized criticality, in contrast, are defined
on a Euklidean neighborhood and any information about robustness
of the system is contained in the excitations of the system. Thus the
dynamical variables fulfill two purposes: On the one hand they act as
the dynamical variables of the system, however, in addition they store
the memory of the self-organized state, for example in terms of a
particular spatial pattern of excitations. One may question, whether
this form of a memory provides a sufficient basis for robustness
as some minimal long term stability is required. Indeed, in some
natural information processing systems one observes that robustness
is stored in topology and details of the local interaction structure
\cite{robustness}.
In this paper we will consider a toy model for self-organization of
communication networks that separates the two time scales of dynamics
and adaptation of the system and also the associated variables.
Models of network evolution have been considered earlier from the
perspective of self-organization \cite{christensen:sneppen:1998}.
Self-organization through coupling of locally adapting links to
an order parameter of the global dynamical phase has shown
to be robust \cite{bornholdt:rohlf:2000}.
Here we wish to address the question of self-organizing networks
for an even simpler system: percolation on a graph with adaptive links.
\newline
Percolation has been a valuable concept for questions of signal
propagation through structures \cite{stauffer:aharony:englishbook}.
Critical properties of a dynamical system can often be conveniently
accessed by observing percolation of damage or perturbations
through the network \cite{kauffman:1969}. One sign of criticality often
observed in dynamical systems is marginal damage spreading across
the system \cite{derrida:pomeau:1986,derrida:1987}. We will study here
how marginal damage spreading can be the result of a dynamical
(adaptation) process itself.
For this purpose let us consider percolation on a directed graph
with the additional property of two types of links, such
that an active site can activate or de-activate a neighbor site,
depending on the specific link. This percolation process is
closely related to the contact process \cite{harris:1974} and the
susceptible-infected-susceptible (SIS) model \cite{SIS}.
In the model studied here, however, de-activation of
active sites does not occur as a decay process with some
given rate, but instead is an interactive process with
an active neighbor site over a de-activating link.
Both, activation and
de-activation of a site thus are regulatory processes.
This process
exhibits a percolation transition where the control parameter is
the fraction of activating links $\vartheta$.
To introduce an adaptive process in this system, the excitatory or
inhibitory characteristics of links are allowed to adapt on a time scale
that is slow in comparison to the time scale of activity propagation
on the network. We will show that percolation at the critical threshold
occurs naturally in this system, when sites tend to equalize their own
average activity with that of their neighbors by locally adjusting their
incoming links.
\newline
The topology chosen in this paper is motivated from immunology,
however, the mechanism is more general and can easily be applied to
systems with different topologies and other motivations.
Here we consider a graph motivated by idiotypic interactions of the immune
system 
\cite{farmer:perelson:1986,valera:coutinho:1991,perelson:weisbuch:1997}.
The graph is spanned by a binary sequence space, where each sequence is
a node representing a possible motif of an immune cell receptor.
Interactions are represented as links: Each motif interacts with its 
complement
and the nearest neighbors of its complement.
Topological properties of and percolation on this so-called one-mismatch 
graph
structure has been discussed in \cite{brede:behn:2001}. A number of 
cellular
automata models to study possible dynamical and functional roles of such
networks in the immune system have been performed (for reviews see
\cite{perelson:weisbuch:1997} and \cite{zorzenon:1999}).
While from an experimental standpoint idiotypic networks are an open
issue today, (neither their large scale structure is known nor a possible
functional role confirmed), from a theoretical view it can be inspiring to
nevertheless speculate about possible functions. We here consider
a possible role in the regulation of activity levels of autoreactive 
lymphocytes
which are present in healthy individuals \cite{ermann:fathman:2001} but
usually do not lead to auto-immune diseases as long as adequate
regulation takes place \cite{goodnow:1996,sherer:shoenfeld:2000}.
At several stages, the immune system shows a preference
for lymphocytes with medium activity \cite{goodnow:1996},
a state that appears to be actively regulated.
One observes that lymphocytes, on the one hand, need some minimal
stimulus for proliferation, but on the other hand may die under chronic
stimulation by self-antigens \cite{goodnow:1996}. Such local interactions
between motifs in the immune system are candidates for systemwide
regulatory mechanisms, however, how a globally stable state of cells
with medium average activities is reached in detail is still an open 
question.
In the following we ask how a network of general dynamical elements
may self-organize to intermediate activity levels of its nodes.
Preference for medium activity should not externally be introduced
but result from local dynamics.
We study a basic mechanism where each node tend to adjust its
own average activity to that of its neighbors. An active site that on
average is more (less) active than its neighbors in the network
experiences more inhibition (activation). As we will see, this results
in marginal percolation of activity across the network corresponding
to a critical state of the system with stable, intermediate average
activity of every node.
\newline
Consider a sequence space of binary strings of length $n$.
Each string codes for a specific immune receptor. Interaction is defined
between complementary sequences and their one-bit-mutants
\cite{brede:behn:2001}.
Each site in sequence space is assigned a Boolean variable
which determines whether the site is active or inactive.
Each active sequence can affect its complementary sequence and the
associated one-bit-mutants in terms of activation or inhibition.
This interaction is represented by directed links between the involved
nodes.
The system is initialized with a fraction $\vartheta$ of activating links.
Starting from an initial condition of sites being mostly inactive and a
small number of activated sites, the following dynamical rules are 
iterated.
\begin{enumerate}
\item Choose a random site.
\item If this site is active evaluate the influence of its $n+1$ outlinks
as follows:
\begin{itemize}
\item If the outlink is activating, the target sequence gets activated.
\item If the outlink is inhibiting, the target sequence gets de-activated.
\end{itemize}
\end{enumerate}
\begin{figure}[b]
\centerline{   \includegraphics[width=6.5cm]{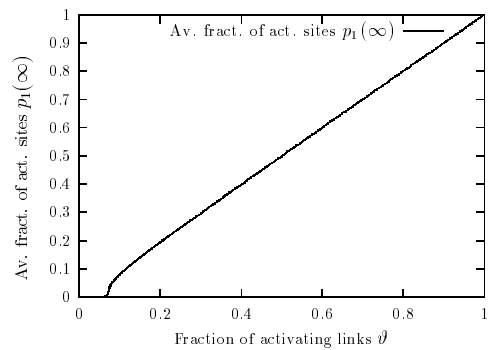}}
\caption{\label{percol}Average fraction of active sites in equilibrium vs.
fraction of activating links $\vartheta$, $n=15$,
$\vartheta^c=\frac{1}{16}=0.0625$}
\end{figure}
As can be seen from figure \ref{percol},
the system shows a percolation transition under variation of $\vartheta$.
The percolation condition is given by
\begin{equation}
\vartheta>\frac{1}{n+1}.
\end{equation}
This corresponds to the condition that each site on average activates
one other site. Above the percolation threshold, the system's equilibrium
state is insensitive to the fraction of active sites at initialization.
The system's evolution can be described by the master equations
\begin{eqnarray*}
\frac{dp_0}{dt}&=&w_{01}p_1-w_{10}p_0\\
\frac{dp_1}{dt}&=&w_{10}p_0-w_{01}p_1
=w_{10}-(w_{10}+w_{01})p_1,
\end{eqnarray*}
where $p_1$ ($p_0=1-p_1$) denotes the probability to find an activated
(inactivated) site and $w_{01}$ and $w_{10}$ represent the respective
transition probabilities. In the mean field limit we can approximate
\begin{eqnarray*}
w_{10}\sim p_1\vartheta, \qquad
w_{01}\sim p_1(1-\vartheta),\\
\frac{dp_1}{dt}\sim p_1\vartheta(1-p_1)-p_1(1-\vartheta)p_1
=p_1\vartheta-p_1^2.
\end{eqnarray*}
Assuming to have reached equilibrium, we demand $\frac{dp_1}{dt}=0$
and get
\begin{equation}
p_1^{\ast}=p_1^{supercrit.}(\infty)=\vartheta.
\end{equation}
Sufficiently far above the percolation threshold the fraction of active
sites after equilibration of the system is identical to the fraction of
activating links.
Below the percolation threshold the fraction of active sites vanishes
for infinite system size. For finite systems, the number of activated
sites in equilibrium may also depend on the initial set of active sites
(cp. figure \ref{percol}).
\newline
Next let us add a slow, local adaptation of links to the model
by which nodes adjust their respective average activity to each other.
We will see that this may drive the system to and stabilize it at the
critical percolation threshold. Define the mean activity $\langle
a_i\rangle_T$ of the site $i$ over some time interval $T$ as
\begin{equation}\label{meanact}
\langle a_i\rangle_T=\frac{1}{T}\sum_{j=1}^T a_i(j),
\end{equation}
with $a_i(j)\in\{0,1\}$ denoting the activity status of site $i$ at time
$j$.
The mean activity of the nearest neighbors of a site $i$ is given by
\begin{equation}\label{meannnact}
\langle a_i\rangle_T^{nn}=\frac{1}{n+1}\sum_{j \in nn}\langle a_j\rangle_T.
\end{equation}
The self-organization of the network takes place by adjusting
the average activity of active sites towards their neighbors' mean
activity, according to the following rules:
\begin{enumerate}
\item Choose a random site $i$.
\item{Spread the activity as in the topologically static model.}
\item If the site is active adjust activating/inhibiting inlinks as
follows:
\begin{itemize}
\item if $\langle a_i\rangle_T>\langle a_i\rangle_T^{nn}$\\
and if the site $i$ has activating inlinks, then reset a random
one of them to be inhibiting with probability $p$.
\item if $\langle a_i\rangle_T\leq\langle a_i\rangle_T^{nn}$\\
and if the site $i$ has inhibiting inlinks, then reset a random
one of them to be activating with probability $p$.
\end{itemize}
\end{enumerate}
These rules are then iterated. For a full run, one first initializes
the system with a supercritical fraction of activating links and
sets all sites to be activated. One then lets the system evolve
as in the static model until it reaches its equilibrium fraction
of active sites $p_1^{\ast} \approx \vartheta$. At that point,
also topological evolution as defined by the above rules
is switched on. We observe that the system equilibrates at a
fraction of activating links that is near the critical value,
independent from the initial fraction of activating links $\vartheta$.
No fine tuning is needed to reach this point. In particular, we
observe that the time scale of equilibration $T$ can be varied
in a wide parameter range without major effects on the limit
fraction of activating links. Furthermore, the convergence is
robust against changes in the probability of local adjustment $p$
which only affects the time scale of equilibration.
To estimate the effect of noise on equilibration, we studied a
variant where, at any iteration, we flip the status of a random link
with some fixed probability $p_{noise}$. One observes that low noise
levels help avoiding suboptimal freezing (sometimes occurring in small
systems) and help approaching the critical fraction of activating links.
Further increasing noise levels, the system equilibrates at higher
than critical levels of activating links.
\newline
An interesting observable of a system at criticality is the size of
dynamical avalanches following small perturbations of the system.
In order to obtain the distribution of avalanches in a system that has
been equilibrated following the above dynamical rules, all
sites are reset to be inactive and then a single random site is activated
and the subsequent activity avalanche is measured. A statistics of
measured avalanche sizes is shown in figure \ref{bothavlb}.
\newline
Avalanche sizes are shown as time-averaged number of active
sites in an avalanche ($\bullet$) and as number of sites that have
been active at some point during an avalanche ($\circ$).
We observe power law scaling with exponents $-1.43$ and $-1.33$,
respectively. The small peak at the right end of the distributions is a
contribution from the giant component of the network, indicating that
the network of this simulation is slightly overcritical. The activating 
links
still form a system-spanning subnetwork here, allowing for broad
spreading of activity.
\begin{figure}[h]
\includegraphics[width=7cm]{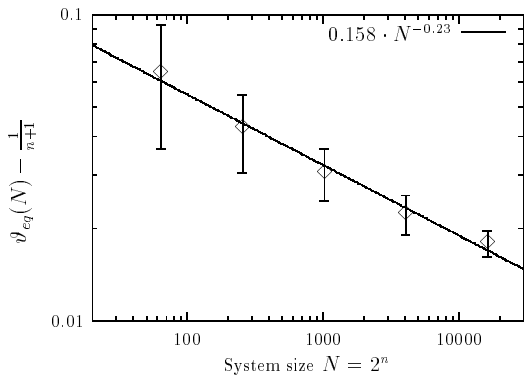}
\caption{\label{fss} Finite size scaling of the fraction of activating
links
$\vartheta_{eq}(N)$, $\vartheta_{eq}(N)-\frac{1}{n+1}$ vs. system size $N=
2^n$,
$\vartheta_{eq}(N)$ approaches $\vartheta^c=\frac{1}{n+1}$ with a scaling
law
$\sim N^{-\beta}$, $\beta=0.23$.}
\end{figure}
Finally, let us study whether the critical point is approached in the
thermodynamic limit using finite size scaling arguments.
Figure \ref{fss} shows the deviation of the equilibrium fraction of
activating links $\vartheta_{eq}(N)$ from the critical value $\vartheta^c$
with increasing system size $N=2^n$. One finds a finite size scaling
relationship for $\vartheta_{eq}(N)$ given by
\begin{eqnarray*}
\vartheta_{eq}(N)-\vartheta^c=\vartheta_{eq}(2^n)-\frac{1}{n+1}=\alpha N^
{-\beta}.
\end{eqnarray*}
Thus we see that the fraction of activating links $\vartheta_{eq}(N)$ of
the equilibrated evolving network converges with increasing system
size $N=2^n$ (as computationally accessible) towards its critical value.
\newline
In this paper we studied percolation on a directed graph with activating
and de-activating links, exhibiting a percolation transition at a critical
fraction of activating links. In an evolution version of this model, links
are allowed to evolve (change their character between activating and
de-activating) on a slow timescale in a way that the average activities of
the two nodes adjacent to the link become more similar to each other.
One finds that this locally defined and 
extremely simplistic rule suffices to
self-organize the network to the percolation transition with high
precision yet without any need for fine tuning.
 From a damage spreading perspective, the condition for critical
percolation is that activity spreads marginally across the system
\cite{kauffman:1969,derrida:pomeau:1986,derrida:1987}.
Equalizing neighboring nodes' activities here shows to induce 
and stabilize marginal damage spreading.
\newline
The particular graph structure used here is motivated by idiotypic immune
networks and the need for balancing the activity of immune cells to an
intermediate level for a functioning immune system. While today one can
at most speculate about possible mechanisms for global immune regulation,
toy models can give important information about dynamical phenomena
that may occur in large complex dynamical systems.
In particular, emergent phenomena
as the one studied in this paper based on simple regulative processes on
the microscale can generate unexpected global behavior that could
manifest itself for example as powerful global regulation of the system.
\newline
The model studied here is not restricted to the particular graph
we used and it would be interesting to study networks with different
topologies as, for example, random graphs or directed networks with
a scale-free degree distribution  \cite{schwartz:havlin:2002}.
Further, it is not difficult to conceive of other models where the basic
principle discussed here, local activity adjustment between neighbors,
could be at work in different contexts. Possible extensions include
neural networks or more complicated cellular automaton models as used,
e.g., in models of social systems. Also extentions of classical SOC
models are possible, such that key variables for self-organization
are stored in separate, more robust variables (as weights or couplings,
for example).
\vspace*{0.3cm}

The authors thank T.\ Rohlf for useful comments on the manuscript.
C.\ Kamp would like to thank the Stiftung der Deutschen Wirtschaft for
financial support.


\begin{thebibliography}{10}

\bibitem{bak:wiesenfeld:1987}
P.~Bak, C.~Tang, and K.~Wiesenfeld,
\newblock Phys. Rev. Lett. {\bf 59}, 381 (1987).

\bibitem{bak:sneppen:1993}
P.~Bak and K.~Sneppen,
\newblock Phys. Rev. Lett. {\bf 71}, 4083 (1993).

\bibitem{drossel:schwabl:1992}
B.~Drossel and F.~Schwabl,
\newblock Phys. Rev. Lett. {\bf 69}, 1629 (1992).

\bibitem{olami:christensen:1992}
Z.~Olami, H.~S. Feder, and K.~Christensen,
\newblock Phys. Rev. Lett. {\bf 68}, 1244 (1992).

\bibitem{robustness}
N.\ Barkai and S.\ Leibler,
Robustness in simple biochemical networks,
Nature 387 (1997) 913-917.

\bibitem{christensen:sneppen:1998}
K.~Christensen, R.~Donangelo, B.~Koiller, and K.~Sneppen,
\newblock Phys. Rev. Lett. {\bf 81}, 2380 (1998).

\bibitem{bornholdt:rohlf:2000}
S.~Bornholdt and T.~Rohlf,
\newblock Phys. Rev. Lett. {\bf 84}, 6114 (2000).

\bibitem{stauffer:aharony:englishbook}
D.~Stauffer and A.~Aharony,
\newblock {\em Introduction to Percolation Theory},
\newblock Taylor and Francis, London, 1992.

\bibitem{kauffman:1969}
S.A.\ Kauffman,
\newblock J.\ Theor.\ Biol.\ {\bf 22}, 437 (1969).

\bibitem{derrida:pomeau:1986}
B.~Derrida and Y.~Pomeau,
\newblock Europhys.\ Lett.\ {\bf 1}, 45 (1986).

\bibitem{derrida:1987}
B.\ Derrida,
\newblock J.\ Phys.\ A: Math.\ Gen.\ {\bf 20}, L721 (1987).

\bibitem{harris:1974}
T.E.\ Harris,
\newblock Ann.\ Prob.\ {\bf 2}, 969 (1974).

\bibitem{SIS}
O.\ Diekmann and J.\ Heesterbeek,
\newblock
Mathematical epidemiology of infectious diseases:
model building, analysis, and interpretation
(John Wiley \& Sons, New York, 2000).

\bibitem{perelson:weisbuch:1997}
A.~Perelson and G.~Weisbuch,
\newblock Rev. Mod. Phys. {\bf 69}, 1219 (1997).

\bibitem{farmer:perelson:1986}
J.D.\ Farmer, N.H.\ Packard, and A.S.\ Perelson,
\newblock Physica D {\bf 22}, 187 (1986).

\bibitem{valera:coutinho:1991}
F.~Valera and A.~Coutinho,
\newblock Immunology Today {\bf 12}, 159 (1991).

\bibitem{brede:behn:2001}
M.~Brede and U.~Behn,
\newblock Phys. Rev. E {\bf 64}, 011908 (2001).

\bibitem{zorzenon:1999}
R.~Zorzenon~dos Santos,
\newblock {\em Immune Responses: Getting Close to Experimental Results with
   Cellular Automata Models}, in: D.\ Stauffer (Ed.),
{\em Ann. Rev. Comp. Phys.}, Vol.\ {VI},
\newblock World Scientific, 1999.

\bibitem{ermann:fathman:2001}
J.~Ermann and C.~Fathman,
\newblock Nature Immunology {\bf 2}, 759 (2001).

\bibitem{goodnow:1996}
C.~Goodnow,
\newblock Proc. Natl. Acad. Sci. USA {\bf 93}, 2264 (1996).

\bibitem{sherer:shoenfeld:2000}
Y.~Sherer and Y.~Shoenfeld,
\newblock Appl. Biochem. Biotechnol. {\bf 83}, 155 (2000).

\bibitem{schwartz:havlin:2002}
N.~Schwartz, R.~Cohen, D.~ben Avraham, A.-L. Barab{\'{a}}si, and S.~Havlin,
\newblock Phys.\ Rev.\ E 64 (2002) 015104.

\end{thebibliography}
\end{document}